# The Agentic Automation Canvas: a structured framework for agentic AI project design

## Authors

- **Sebastian Lobentanzer** ✉
  [https://orcid.org/0000-0003-3399-6695](https://orcid.org/0000-0003-3399-6695) · [slobentanzer](#)
  Institute of Computational Biology, Computational Health Center, Helmholtz Center, Munich, Germany; German Center for Diabetes Research, Munich, Germany; Open Targets, European Bioinformatics Institute (EMBL-EBI), Hinxton, Cambridge, United Kingdom

✉ — Correspondence possible via [GitHub Issues](#) or email to Sebastian Lobentanzer <sebastian.lobentanzer@helmholtz-munich.de>.

## Abstract

Agentic AI prototypes are being deployed across domains with increasing speed, yet no methodology for their structured design, governance, and prospective evaluation has been established. Existing AI documentation practices and guidelines – Model Cards, Datasheets, or NIST AI RMF – are either retrospective or lack machine-readability and interoperability. We present the Agentic Automation Canvas (AAC), a structured framework for the prospective design of agentic systems and a tool to facilitate communication between their users and developers. The AAC captures six dimensions of an automation project: definition and scope; user expectations with quantified benefit metrics; developer feasibility assessments; governance staging; data access and sensitivity; and outcomes. The framework is implemented as a semantic web-compatible metadata schema with controlled vocabulary and mappings to established ontologies such as Schema.org and W3C DCAT. It is made accessible through a privacy-preserving, fully client-side web application with real-time validation. Completed canvases export as FAIR-compliant RO-Crates, yielding versioned, shareable, and machine-interoperable project contracts between users and developers. We describe the schema design, benefit quantification model, and prospective application to diverse use cases from research, clinical, and institutional settings. The AAC and its web application are available as open-source code and interactive web form at [https://aac.slolab.ai](https://aac.slolab.ai).

# Main

Agentic AI systems—autonomous software agents that can plan, reason, and execute multi-step tasks with minimal human oversight—are rapidly emerging across all fields of science [1,2]. From automated literature curation and clinical data extraction to autonomous laboratory experimentation, these systems promise transformative gains in efficiency, quality, and scalability. Yet they also introduce a fundamental shift in how work is organized: unlike traditional software tools, agentic systems require an inversion of the operational control loop; the system assumes command of tasks while humans step back from moment-to-moment decision-making [1,2]. This shift necessitates new forms of negotiation between stakeholders, explicit governance structures, and documentation practices that go well beyond what conventional software development processes provide. In the current landscape, there is no established planning process for agentic systems; they are conceived *ad hoc*, and they are evaluated *ad hoc*. Often, the balance between user expectations and technical feasibility is not clear, and the record of project details and governance is neither standardised nor machine-readable. This can lead to the surprising finding that these systems are much less effective in practice than expected, displaying an *expectation–realisation gap* [3,4,5].

We present the Agentic Automation Canvas (AAC), a structured framework for designing, governing, and documenting agentic automation projects. Inspired by the Business Model Canvas [6], which provides a single-page structured overview of a business model's key components, the AAC captures six interconnected dimensions of an agentic automation project: project definition, user expectations, developer feasibility, governance staging, data access and sensitivity, and outcomes (Figure 1 a).

The central innovation is the formalization of a bidirectional *contract* between users and developers. User expectations are captured as structured requirements with quantified benefit metrics across five dimensions: time savings, quality improvement, risk reduction, enablement of new capabilities, and cost efficiency. Developer feasibility assessments evaluate the technical reality of delivering these benefits, including technology readiness levels, model selection, implementation architecture, and risks (Figure 1 b). The canvas is primarily designed as a communication tool; ideally, users and developers fill it out together (Figure 1 c). By requiring both perspectives, the AAC surfaces misalignments early—before significant resources are committed.

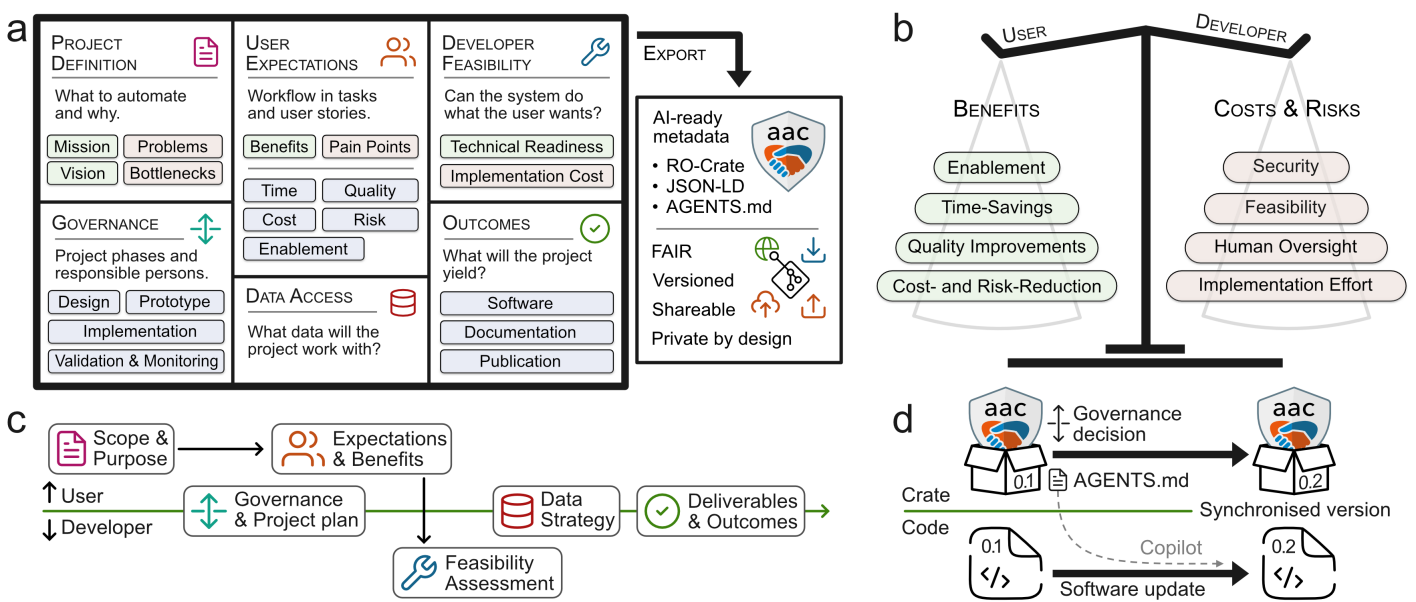

**Figure 1: The Agentic Automation Canvas. a)** Overview of the six canvas dimensions. Single projects are captured in FAIR, versioned, and shareable RO-crates. By being fully client-side, the application is private by design. **b)** The canvas is a communication tool to be used between users and developers of a project, balancing expected gains against technical feasibility considerations. **c)** The communication process between users and developers reflects the dimensions of the canvas and the timeline of project development. The canvas helps structure this conversation. **d)** Synchronising the versioning of the canvas RO-Crate as a living document and the implementation codebase as the real-world instance of the plan facilitates project management and reporting. The integrated AGENTS.md description can facilitate implementation using copilots or coding agents.

The AAC is implemented as a fully semantic web-compatible metadata schema, made accessible by an interactive web application that guides users through structured data capture with real-time validation (https://aac.slolab.ai). Crucially, the application runs entirely in the browser—there is no server-side processing, and canvas data never leaves the user's machine. While agentic automation canvases for clinical workflows, proprietary pipelines, or sensitive research programs may contain information that cannot be shared with external services, the machine-interoperable format facilitates transparent public sharing and aggregation, if desired.

Canvases can be exported at any time as Research Object Crate (RO-Crate) packages [7], self-contained digital objects that bundle the canvas specification with standards-compliant metadata following established ontologies such as Schema.org [8] and W3C DCAT [9] (see Methods for details). The resulting packages are FAIR compliant [10] but under the user's full control: they can remain strictly internal, be shared selectively with collaborators or governance boards, or be published openly in research data repositories. They are also AI-ready: the RO-Crate export includes an AGENTS.md file that translates the design contract into instructions consumable by coding copilots and LLM-based development agents (Figure 1 d). For iteration, canvases can simply be re-uploaded to the interactive web form. Canvases carry semantic version numbers, allowing teams to track how a project's design evolves over time and to synchronize the living document with actual implementation milestones (Figure 1 d). The additional classes we define in the AAC schema offer a structured vocabulary for the rapidly evolving landscape of agentic AI implementation patterns, complementary to the existing semantic web base vocabulary. The schema profile is defined independently of the web application (https://w3id.org/aac), enabling development of command-line tools, API integrations, and programmatic workflows.

We are applying the AAC to numerous projects spanning single-cell bioinformatics, clinical research assistants, drug target databases, patient-facing chatbots, research data management, and institutional AI coordination. Use cases range from individual project design through portfolio management to research infrastructure specification. For explanation, we give an example of making the Open Targets drug discovery platform more accessible through agentic natural language interfaces (see also the RO-Crate in Supplementary File SF1). The users are drug discovery experts with biological background who want to query drug-target associations conversationally rather than through complex database interfaces. Expected benefits are quantified as time-to-insight reduction and accessibility for non-computational researchers. Developer feasibility assesses that retrieval-augmented generation using a custom MCP server over the Open Targets knowledge graph is technically viable but carries medium-high risk, particularly for queries spanning proprietary partner data (restricted access) and public evidence (open). Governance staging defines validation by the platform team during prototyping, with staged rollout from internal testing to public deployment. Outcome metrics link back to the initial benefit estimates, enabling direct comparison of expected versus actual gains. Without the canvas, these considerations would live in scattered documents and slide decks—if captured at all.

This combination of prospective design, quantified contracts, and FAIR-compliant outputs distinguishes the AAC from existing approaches to AI documentation and governance (Table 1). Model Cards [11] and Data Cards [12] document trained models and datasets retrospectively. The Machine Learning Canvas [13] adapts the canvas format to ML project design but focuses on product-market fit rather than governance, benefit quantification, or machine-readable specifications. AI governance frameworks such as the NIST AI RMF [14] provide compliance checklists but not machine-readable specifications that integrate with data management infrastructure. The AAC is prospective: it guides design decisions before and during development, when they can still influence outcomes, and produces machine-readable RO-Crate exports that encode benefit models, risk assessments, and governance staging. Policy Cards [15] complement this landscape by defining a deployment-layer, normative specification of operational constraints for deployed AI agents. Where the AAC captures planning intent, Policy Cards enforce runtime constraints; together they span the full lifecycle from design through deployment to audit (for more detail, see Methods).

**Table 1:** Positioning of AI documentation and governance artifacts across the system lifecycle. The AAC fills the prospective planning stage with machine-readable specifications that can feed into normative Policy Cards at deployment. Adapted from [15].

| Artifact | Primary Scope | Lifecycle Stage | Normativity | Machine-Readable Audit Hooks |
|---|---|---|---|---|
| **Data Card** [12] | Dataset provenance | Post-collection | Descriptive | Limited (links/IDs) |
| **Model Card** [11] | Trained model reporting | Post-training | Descriptive | Limited (links/IDs) |
| **AAC (this work)** | Project design, benefits, risks, governance | Pre-development through deployment | Prospective/Communicative | Yes (RO-Crate, benefit models, risk registers, compliance mappings) |
| **Policy Card** [15] | Deployment policy for a specific system/context | Deployment/Runtime | Normative (allow/deny/escalate) | Yes (rules, evidence, thresholds, assurance mappings) |

The benefit quantification model provides a structured language for the value proposition of agentic automation. The AAC requires explicit, measurable expectations with human oversight costs factored in, enabling realistic assessment of net benefits. This structured approach to value estimation supports informed go/no-go decisions about whether to pursue, continue, or terminate agentic automation initiatives. The RO-Crate output format ensures that these specifications persist as reusable digital objects, enabling cross-project comparison, institutional learning, and community development of best practices around agentic system design.

We implement our diverse use cases as prospective case studies with explicit pre-registration of expectations, the analysis of which will allow quantitative insights into the automation of complex processes in the research landscape. However, outside of projects with our direct involvement, a significant limitation of the decentralised private-by-design RO-Crates as the only output of the canvas is the need for additional mechanisms for sharing and finding them. In the future, we aim to integrate the resulting metadata repository more closely with agentic tooling such as our MCP server registry [16], the registry for our knowledge graph framework components [17], and community research software platforms [18], in order to allow the broad reuse and recomposition of open-source agentic workflows.

The canvas is *not* meant as a contract to be *enforced;* rather, it is designed as a living document to facilitate the deliberate, transparent, and accountable conversation between users and developers from different backgrounds. For any given project, the canvas is expected to change through the development process; thus, it is adaptable by design. We welcome contributors and use cases from the community, which we will continually support through open-source mechanisms.

# Methods

## Canvas Data Model

The Agentic Automation Canvas schema defines a structured data model organized into six primary sections.

The **project definition** captures core metadata: title, description, objectives, development stage (planning, prototype, or deployment), domain classification, keywords, funding information, lead organization, and a project-level value summary including headline value statement and primary value driver.

**User expectations** consist of structured requirements, each with a title, description, user story, priority, status, unit of work definition, monthly volume, an optional target population specifying which user group the benefit estimates apply to, and an array of **benefit metrics** (detailed [below](below)). Requirements can declare dependencies on other requirements, enabling workflow modeling. Stakeholders are linked to requirements by referencing centrally managed person entities.

**Developer feasibility** operates at two levels: project-level defaults (technology readiness level, overall technical risk, effort estimate) that apply to all tasks, and optional per-task overrides that capture algorithm specifications, tool requirements, model selection (open-source, frontier-model, fine-tuned, custom, or none), and technology architecture including simple prompting, retrieval-augmented generation (with retrieval method, embedding model, and chunking strategy), fine-tuning (with base model, approach, and dataset), and agentic frameworks (with framework, tools, and orchestration details). Per-task feasibility can link to Model Cards [11] via identifier fields where a trained model is in use and includes structured risk assessments described in detail [below](below).

**Governance staging** defines lifecycle phases with start and end dates, agents responsible for decisions (persons, organizations, or software systems), milestones with KPIs, compliance standards (as plain strings or structured framework references with specific clauses and URIs), and optional Policy Card URIs linking to machine-readable deployment governance artifacts [15].

**Data access and sensitivity** captures dataset metadata including format, license, access rights (open, restricted, confidential, or highly restricted), sensitivity level, personal data indicators, DUO terms for use restrictions, and persistent identifiers, with fields to link to Data Cards [12] where applicable.

**Outcomes** track deliverables (with type, status, and persistent identifiers), publications (with DOIs and author lists), and evaluation results (with metrics, methods, and findings).

All persons involved in a project are managed in a centralized persons registry with unique identifiers, names, affiliations, ORCID identifiers, and functional roles from a controlled vocabulary, enabling consistent identity management and role aggregation across the canvas.

## Benefit Quantification Model and Risk Assessment

The AAC captures expected benefits through a generalized benefit structure that supports five types: **time**, **quality**, **risk**, **enablement**, and **cost**. Benefits are always positive: they represent expected gains (e.g. time saved, quality improved, risk reduced, new capabilities enabled, cost reduced). The same qualities can, however, also affect the risk profile—for example, expected time savings may be undermined by technical delays, or expected quality gains by data-quality issues—so the canvas pairs benefit quantification with structured risk assessment.

Each **benefit metric** specifies a metric identifier (from a controlled vocabulary or custom), a human-readable label, a direction indicating whether higher, lower, target, or boolean values are preferred, and whether values represent absolute measures or deltas. Baseline and expected values can be numeric, categorical (low/medium/high), or binary, accommodating diverse measurement approaches.

For time benefits, human oversight values (minutes per unit of work or minutes per month) are explicitly captured and subtracted from gross time savings, producing a realistic net benefit estimate. Each benefit includes an aggregation basis (per unit, per month, or one-off), confidence levels from both user and developer perspectives (low/medium/high), and free-text assumptions documentation. Benefits aggregate at the project level through volume-weighted calculations, providing headline metrics for decision-making.

Per-task **risk assessments** use six categories that parallel and complement the benefit dimensions and are recorded as structured records. Each risk specifies a category (technical, data, compliance, operational, ethical, or adoption), a title, a free-text description, likelihood and impact ratings (low through critical), a mitigation strategy, and a status (identified, mitigated, accepted, or resolved). **Technical** risks (e.g. model failure, integration issues) can threaten time and cost benefits and undermine delivery. **Data** risks (e.g. quality, availability, access) directly parallel quality benefits and can limit enablement. **Compliance** risks (e.g. regulatory breach) oppose the benefit type "risk" (risk reduction) and can create legal or reputational harm. **Operational** risks (e.g. reliability, support) affect time, cost, and sustained quality. **Ethical** risks (e.g. bias, fairness, misuse) can negate quality and risk-reduction gains and trigger governance obligations. **Adoption** risks (e.g. uptake, change management) determine whether enablement and related benefits materialise in practice. By pairing benefit types with these risk categories, the canvas supports a balanced view of both upside and downside for each requirement.

## Standards Compliance

The AAC generates RO-Crate 1.2 packages that adhere to the following standards and ontologies:

**RO-Crate 1.2** [7]: The Research Object Crate specification provides a standardized packaging format for research outputs with their metadata. Each AAC export produces a ZIP archive containing an `ro-crate-metadata.json` file (JSON-LD), a human-readable preview (`ro-crate-preview.html`), the original canvas data (`canvas.json`), and documentation files.

**Schema.org** [8]: Project metadata is structured using Schema.org types including `Project`, `ResearchProject`, and `CreativeWork`, enabling discovery through web search engines and metadata catalogs. Persons are typed as `schema:Person` with `schema:name`, `schema:affiliation`, and `schema:identifier` (ORCID) properties.

**W3C DCAT** [9]: Dataset metadata follows the Data Catalog Vocabulary, with each dataset represented as `dcat:Dataset` including `dcat:distribution`, `dcat:accessRights`, and `dcat:contactPoint` properties, enabling integration with data catalogs.

**W3C PROV-O** [19]: Governance activities and their relationships are captured using the Provenance Ontology. Governance stages are modeled as `prov:Activity` instances with `prov:wasAssociatedWith` linking to agents and `prov:generated` linking to milestones.

**P-Plan** [19]: User expectations and requirements are structured using the Plan Ontology, with each requirement represented as a `p-plan:Step` within a `p-plan:Plan`, and dependencies modeled as `p-plan:isPreceededBy` relationships.

**FRAPO** [20]: Funding and project administration information follows the Funding, Research Administration & Projects Ontology, enabling integration with research administration systems through `frapo:Grant` and `frapo:FundingAgency` entities.

**DUO** [21]: Data use restrictions are specified using controlled terms from the Data Use Ontology, enabling automated compliance checking. Terms such as `DUO:0000006` (health or medical or biomedical research) and `DUO:0000007` (disease-specific research) provide machine-readable access conditions.

## Schema Profile

The AAC schema profile is maintained independently of the web application at https://w3id.org/aac/ and includes the following components:

**JSON Schema** (`canvas-schema.json`): A formal JSON Schema (Draft 07) specification that validates canvas data structure. The schema enforces required fields (project title, description, and stage), validates enumerated values (technology readiness levels, risk levels, `DUO` terms, benefit types, access rights), and ensures referential integrity between person identifiers and their references in stakeholder and agent roles.

**RO-Crate Profile** (`rocrate-profile.json`): Defines the expected structure of generated RO-Crate packages, including required entity types, properties, and relationships, enabling validation of exported packages against the profile.

**Controlled Vocabularies**: Standardized term lists for technology readiness levels (1–9), `DUO` terms, governance stages, risk levels (low, medium, high, critical), and functional roles, distributed as JSON files within the schema package.

**Ontology Mappings**: Detailed documentation of how canvas concepts map to each ontology, provided as human-readable Markdown files covering Schema.org, DCAT, PROV-O, P-Plan, FRAPO, and `DUO` mappings.

## Implementation

The web application is built with Vue.js 3 (`v3.5.27`), TypeScript (`v5.9.3`), Vite (`v7.3.1`), and Tailwind CSS (`v3.4.19`). The version of the schema and application at the time of this publication is `v0.13.1`. Dependencies are managed with **npm** for the front-end (Node.js) and **uv** for the documentation and tooling environment (Python); the exact resolved versions are recorded in the repository lockfiles (package-lock.json, uv.lock). The interface is organized into sections corresponding to the six canvas dimensions, with collapsible panels, contextual help tooltips, and form validation providing immediate feedback on data completeness and correctness. Complex data structures (requirements, stakeholders, governance stages, datasets) support add, edit, and delete operations with nested sub-forms.

The RO-Crate generation pipeline validates canvas data against the JSON Schema specification, transforms it into RO-Crate-compliant JSON-LD using the ontology mappings, generates an HTML preview, and packages all files into a ZIP archive. Users can import existing canvas JSON files for iterative editing and template reuse. The application is deployed as a static site, requiring no server-side infrastructure, and is accessible at https://aac.slolab.ai.

## From Planning to Deployment: AAC-to-Policy-Card Mapping

The AAC occupies the planning phase of the agentic system lifecycle, while Policy Cards [15] occupy the deployment and runtime phase. Policy Cards are machine-readable, normative specifications that define what a deployed agent must and must not do, including allow/deny/escalation rules (ABAC-style controls), monitoring requirements, KPI thresholds with auto-fail conditions, and assurance mappings to frameworks such as NIST AI RMF, ISO/IEC 42001, and the EU AI Act. These two artifacts are complementary: the AAC captures the planning intent, and the Policy Card formalises the deployment constraints.

We propose a structured mapping from AAC canvases to Policy Cards that can be implemented as a standalone transformation tool operating on the RO-Crate export. The mapping follows the natural information flow from planning to deployment (Figure 2):

- **Governance compliance standards** (`governance.stages[].complianceStandards`) map to `applicable_policies` and `assurance_mapping` in the Policy Card, with structured framework references (framework name, specific clauses, URIs) translating directly to assurance tokens.
- **Per-task risk assessments** (`requirements[].feasibility.risks`) inform `controls.action_rules`: identified risks with high likelihood or critical impact translate to deny or require_escalation rules, while mitigated risks may translate to allow rules with evidence requirements.
- **Governance agents** (`governance.stages[].agents`) map to `scope.stakeholders`, preserving the person/organisation/software typing.
- **Dataset access rights** (`dataAccess.datasets[].accessRights` and `DUO` terms) generate data-related controls, such as deny rules for processing highly restricted data without appropriate authorisation.
- **Benefit KPIs** (`requirements[].benefits`) inform `kpis_thresholds`, with baseline and expected values translating to operational target thresholds, and critical benefits potentially generating `critical_auto_fail` conditions.
- **Policy Card URIs** (`governance.stages[].policyCardUri`) provide a direct reference link, enabling round-trip traceability between the planning artifact and the deployment artifact.

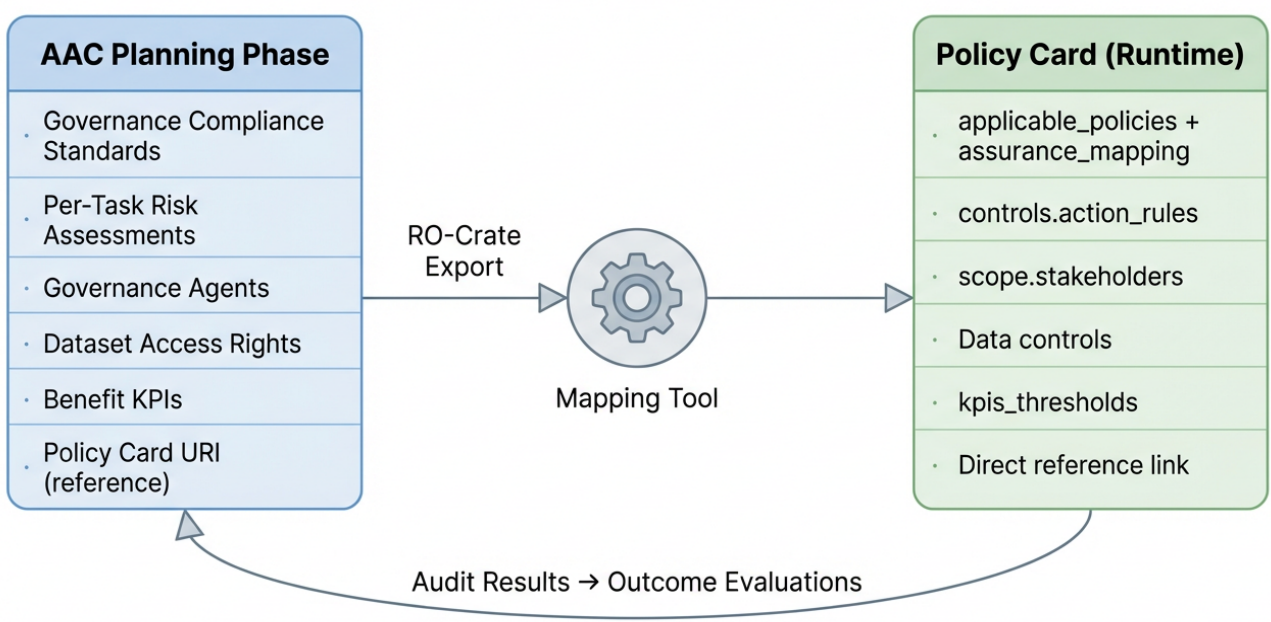

**Figure 2: Proposed AAC-to-Policy-Card mapping workflow.** The AAC canvas (left) captures planning-phase intent across its six dimensions. An automated mapping tool reads the exported RO-Crate and transforms relevant fields into a Policy Card (right), which governs the deployed agent at runtime. Policy Card audit results feed back into AAC outcome evaluations, closing the governance loop.

The transformation tool itself is future work. We anticipate that it will read the AAC RO-Crate, apply the field mappings described above, and generate a baseline Policy Card JSON document that can be reviewed, refined, and deployed. This workflow treats the AAC as the "Declare" phase input for the Policy Card's Declare-Do-Audit lifecycle, establishing a continuous governance chain from project planning through runtime enforcement to post-deployment assurance.

# Code Availability

The Agentic Automation Canvas web application and schema are openly available under the Apache License 2.0. Source code is maintained at [https://github.com/slolab/agentic-automation-canvas](https://github.com/slolab/agentic-automation-canvas). A citable archive of all versions of the software is available from Zenodo at [https://doi.org/10.5281/zenodo.18649597](https://doi.org/10.5281/zenodo.18649597). The schema profile is available under the persistent W3ID namespace [https://w3id.org/aac](https://w3id.org/aac).